\documentclass[a4paper]{article}

\usepackage{16lomcon}        
\usepackage{cite}             
\usepackage{epsfig}           

\def\roughly#1{\mathrel{\raise.3ex\hbox
{$#1$\kern-.75em\lower1ex\hbox{$\sim$}}}}

\def\gs{\roughly>}

\begin{document}


\title{PLASMON DECAY TO A NEUTRINO PAIR VIA NEUTRINO ELECTROMAGNETIC MOMENTS IN A STRONGLY MAGNETIZED MEDIUM}

\author{A. V. Borisov \email{borisov@phys.msu.ru}}

\affiliation{Faculty of Physics, M. V. Lomonosov Moscow State University,\\ 119991
Moscow, Russia}

\author{P. E. Sizin \email{mstranger@list.ru}}

\affiliation{Department of Higher Mathematics, Moscow State Mining University,\\119991 Moscow, Russia}

\date{}
\maketitle


\begin{abstract}
We calculate the neutrino luminosity of a degenerate electron gas in
a strong magnetic field via plasmon decay to a neutrino pair due to
neutrino electromagnetic moments and obtain the relative upper
bounds on the effective neutrino magnetic moment.
\end{abstract}


~~~{\bf 1.} Neutrino emission is the main mechanism of the
energy losses of stars in the late stages of their evolution
\cite{r1}. We will consider cooling of the outer regions of
neutron stars that are rarefied enough to be transparent to originating neutrinos. Strong
magnetic fields ($H\gs 10^{12}~{\rm G}$) can exist in these
regions; moreover, the fields for the class of the neutron stars that are called magnetars can reach $10^{14}-–
10^{16}~{\rm G}$ \cite{r2} (see also \cite{r3}).

Under these conditions, the main processes of neutrino production
are annihilation of an electron-positron pair ($e^{-}e^{+}\rightarrow\nu\bar\nu$),
photoproduction of a neutrino pair on
the electron ($\gamma e^{\pm}\rightarrow e^{\pm}\nu\bar\nu$), photon decay ($\gamma\rightarrow\nu\bar\nu$) ,
and two-photon annihilation ($\gamma\gamma\rightarrow\nu\bar\nu$). The
results of the study of these processes (without a magnetic field) were given in the review \cite{r4}. The luminosity
of a degenerate nonrelativistic gas via photoproduction of neutrino pairs for the case of a superstrong
magnetic field was calculated in \cite{r5}. The authors of \cite{r6} estimated the luminosity of the degenerate electron gas
due to these processes in a superstrong field. The results for photoproduction of neutrino pairs were corrected in \cite{r7}.

Simple extension of the standard model of the electroweak interactions generates electromagnetic dipole
moments of a massive Dirac neutrino (see \cite{r1} and a recent review \cite{r8})).

~~~{\bf 2.} In this report, we address one of the processes of neutrino emission that is plasmon decay to a neutrino pair mediated by the neutrino electromagnetic moments.
As is well known, the plasmon is the photon with a nonzero mass generated by interaction with a medium. A relevant medium model for the outer region of the neutron star is
a degenerate electron gas in a strong magnetic field $H$:
\begin{equation}
\label{cond}
T\ll \mu -m,\, H>((\mu/m)^2 - 1)H_0/2,
\end{equation}
where $T$ and $\mu\simeq \mu(T=0)\equiv \varepsilon_{\rm F} = (m^2 + p_{\rm F}^2)^{1/2}$ are the temperature and chemical potential of the gas, $\varepsilon_{\rm F}$ and $p_{\rm F}$ are the Fermi energy and momentum, $H_0 = m^2/e \simeq 4.41\times 10^{13}~{\rm G}$, $m$ and $-e$ are the electron mass and charge (we use the units with $\hbar=c=k_{\rm B}=1$). Under the conditions (\ref{cond}), electrons occupy only the lowest Landau level in the magnetic field with $p_{\rm F} = 2\pi^2 n_e/(eH)$, where $n_e$ is the electron concentration, and the effective photon mass is generated which is equal to the plasmon frequency \cite{r9}
\begin{equation}
\label{pl}
\omega_p = ((2\alpha/\pi)(p_{\rm F}/\varepsilon_{\rm F})H/H_0)^{1/2}m,
\end{equation}
$\alpha$ is the fine-structure constant.

Taking into account that for the relatively small momentum of the photon $k$ in the analyzed process
the vertex operator of the photon–neutrino coupling (for the Dirac neutrino) is as follows \cite{r1,r8}
${\Gamma ^\alpha } = {\sigma ^{\alpha \beta }}{k_\beta }({\mu _\nu } + i{\gamma ^5}{d_\nu })$, we have calculated
the luminosity (the rate of energy losses by a unit volume of a medium) due to the process $\gamma\rightarrow\nu\bar\nu$
through the electromagnetic channel
\begin{equation}
\label{lum}
Q_{\rm em}=\frac{\bar{\mu}_{\nu}^2\omega_p^4}{48\pi^3}
\int\limits_{0}^\infty \frac{k^2 dk}{e^{\frac{1}{T}\sqrt{\omega_{p}^2+k^2}}-1}\,,
\end{equation}
where the effective neutrino magnetic moment $\bar{\mu}_{\nu} = \sqrt{\mu_\nu^2 + d_\nu^2}$.

~~~{\bf 3.} The upper (relative) bound on $\bar{\mu}_{\nu}$ will be
found from the following requirement: the luminosity (\ref{lum}) should be
lower than that in a weak channel $Q_{\rm w}$. Comparing Eq. (\ref{lum}) with the result for $Q_{\rm w}$
from \cite{r6}, we obtain
\begin{equation}
\label{limap}
{\hat \mu _\nu } = \bar{\mu}_{\nu}/{\mu}_{\rm B} < 1.58\times 10^{-12}T_{8}F(p) \ge 3.60\times 10^{-12}T_8.
\end{equation}
Here, ${\mu}_{\rm B}$ is the Bohr magneton, the function $F(p) = p\left[ {\bar g_V^2 + \frac{2}{3}\,\bar g_A^2\frac{{{B_4}(p)}}{{{B_2}(p)}}} \right]^{1/2}$
with ${B_n}(p) = \int\limits_0^\infty  {\frac{{{x^n}dx}}{{\exp (p\sqrt {1 + {x^2}} ) - 1}}}$, the effective weak couplings
${\bar g}_V^2 \simeq 0.929$, ${\bar g}_A^2 = 3/4$, the argument (see Eq. (\ref{pl})) $p = {\omega _p}/{T} =
1.92{\left( {1 + 0.44H_{13}^2\rho _6^{ - 2}} \right)^{ - 1/4}}H_{13}^{1/2}T_8^{ - 1}$ (under the conditions of the neutron star
crust \cite{r3,r4}, the electron density is expressed through the matter density $\rho$ and the proton mass $m_p$: $n_e \simeq 0.5\rho/m_p$), and $H_{13} = H/(10^{13}~\mbox{G})$, $T_8 = T/(10^8~\mbox{K})$, $\rho _6 = \rho /(10^6~\mbox{g}/\mbox{cm}^3)$.

For the case $\omega_p\ll T$ ($p \ll 1$), we obtain from (\ref{limap}) the bound:
\begin{equation}
\label{lim<}
\hat \mu _\nu < 3.6 \times 10^{ - 12}\, T_8\,.
\end{equation}
In particular, at $T_8=1.8$ (as in \cite{r7}) we get the bound $ \hat \mu _\nu < 6.5 \times 10^{ - 12}$,
which is slightly weaker than that found in \cite{r7} ($ \hat \mu _\nu < 1.1 \times 10^{ - 12}$) from the comparison 
of the electromagnetic and weak mechanisms of the photoproduction $\gamma e \to e\nu \bar \nu$, which is more effective
under the same conditions than the plasmon decay \cite{r6}.

For the case $\omega_p\gg T$ and a nonrelativistic electron gas ($p_{\rm F} \ll m$, that is $H_{13}/\rho_6 \gg 1$), from  (\ref{limap}) it follows
\begin{equation}
\label{lim>nr}
 \hat \mu _\nu < 3.61 \times 10^{ - 12}\, \rho_6^{1/2}\,.
\end{equation}

For $\omega_p\gg T$ and $p_{\rm F} \gg m$ (a relativistic gas), we obtain
\begin{equation}
\label{lim>r}
\hat \mu _\nu  < 2.94 \times 10^{ - 12}\, H_{13}^{1/2}\,.
\end{equation}
The analysis shows that the conditions $p_{\rm F} \gg m$ and (\ref{cond})
can be simultaneously satisfied only at a rather strong field
$H$. For example, at $H_{13} = 300$, Eq. (\ref{lim>r}) gives $\hat \mu _\nu  < 5.1 \times 10^{ - 11}$, which
is close to the bounds $\mu_{\nu } < 5.4\times 10^{-11}\mu _{\rm B}$  and $\mu_{\bar\nu_e } < 2.9 \times 10^{-11}\mu _{\rm B}$ that were obtained from the analysis of
solar neutrinos \cite{r10} and in the GEMMA laboratory experiment on antineutrino scattering off electrons \cite{r11}, respectively.

~~~{\bf 4.} Relative upper bounds on the effective neutrino magnetic moment (see Eqs. (\ref{limap})--(\ref{lim>r})) determine
the range of its values where the weak channel of the plasmon decay is more effective than the electromagnetic one. In conclusion, we note that production of a
neutrino pair by a high-energy photon was considered in \cite{r12}. As opposed to the plasmon decay discussed
above, this process is caused by coherent interaction of a neutrino possessing a magnetic moment with a dense
medium.

We thank A.~E.~Lobanov for useful discussions.

\end{document}